\documentclass[aps,prb,twocolumn,superscriptaddress,longbibliography]{revtex4-1}
\usepackage{epsfig,amsopn}
\usepackage{graphicx}
\usepackage{color}
\usepackage{amsmath,amssymb}
\usepackage{enumerate}
\newcommand\bea{\begin{eqnarray}}
\newcommand\eea{\end{eqnarray}}
\newcommand\beq{\begin{equation}}
\newcommand\eeq{\end{equation}}

\def\nn{\nonumber}
\def\f{\frac}

\def\si{\sigma}
\def\Do{\partial}

\def\ra{\rangle}
\def\ua{\uparrow}
\def\da{\downarrow}

\def\th{\theta}

\begin{document}
\title{Transport across junctions of altermagnets with normal metals and ferromagnets} 
\author{Sachchidanand Das}
\affiliation{School of Physics, University of Hyderabad, Prof. C. R. Rao Road, Gachibowli, Hyderabad-500046, India}
\author{Dhavala Suri}
\affiliation{Department of Physics, Technical University of Munich, Garching 85748, Germany}
 \author{ Abhiram Soori~~}
 \email{abhirams@uohyd.ac.in}
 \affiliation{School of Physics, University of Hyderabad, Prof. C. R. Rao Road, Gachibowli, Hyderabad-500046, India}
\begin{abstract}
Altermagnet (AM) is a novel time reversal symmetry broken magnetic phase with $d$-wave order which has been experimentally realized recently. We discuss theoretical models of altermagnet based systems on lattice and in continuum. We show equivalence between the lattice and continuum  models by mapping the respective parameters. We study (i) altermagnet-normal metal (NM) and (ii) altermagnet-ferromagnet (FM) junctions, with the aim to quantify transport properties such as conductivity and  magnetoresistance.  We find that a spin current accompanies charge current when a bias is applied. The magnetoresistance of AM-FM junction switches sign when AM is rotated by $90^{\circ}$ -- a feature unique to the altermagnetic phase. 
\end{abstract}
\maketitle
\section{Introduction}

One of the cornerstones of  spintronics is the class of ferromagnetic  metals (FM), where the net spin polarization in the band structure is exploited to realize a variety of exciting phenomena \cite{wolf01}.  Examples include realization of spin Hall effect in FM   where spins accumulate on edges of the sample~\cite{murakami03,sinova04,shermp}; giant magnetoresistance, which has applications in data storage~\cite{chappert07}; and FMs in combination with spin orbit coupled systems  known as the Datta-Das transistor~\cite{dattadas,chuang2015,Choi2015,bijay22}.   The emergence of  antiferromagnetic spintronics in recent years has added to the plethora of spintronic phenomena that are experimentally relevant~\cite{baltz18,fukami20,hoffmann22}.  Recent advances in spintronics have advanced into yet another phase of magnetism, called the altermagnets \cite{smejkal22a}.   A prominent example of an altermagnet is the Ruthenium dioxide (RuO$_2$), which is found to host altermagnetic ordering~\cite{berlijn17,zhu19,lovesey22}. Other candidate materials are MnO$_2$, MnF$_2$, FeSb$_3$ and CaCrO$_3$ ~\cite{smejkal22b}; RuO$_2$ is metallic in contrast to others which are insulating. The crystal structure of these materials has two spin sub-lattices with opposite spins, resulting in anisotropic Fermi surfaces. In momentum space, the down-spin band is obtained by $90^{\circ}$ rotation of the up-spin band; implying that the hopping strength in the real-space lattice depends on both the direction and spin of electrons. Thus, the  time-reversal symmetry (TRS)~\cite{smejkal20,feng22} is broken, although the net magnetization remains zero. A combination of TRS broken magnetic texture combined with spin-orbit coupling can result in anomalous Hall effect, similar to transverse currents in spin-orbit coupled systems combined with a Zeeman field~\cite{soori2021,bijay22}. Furthermore, spin Hall effect is possible in altermagnets without the need for spin-orbit coupling~\cite{bai22,karube22,bose22}, indicating the prospects of the altermagnetic phase to spintronics.

Magnetism is known to play a critical role in mesoscopic devices to enhance exotic scattering processes in superconducting systems~\cite{yamashita03,beckmann04,soori22car,soori18sar}, in which fundamental excitations  consist of electrons and holes.  A comparison between superconducting and FM states reveals an interesting similarity between electron-hole degrees of freedom in superconductors  and the two spin states of ferromagnets. While the s-wave superconductor is like ferromagnet,  p-wave superconductor is similar to spin-orbit coupled systems. A question then arises~\cite{schofield09} -` what is the equivalent magnetic phase corresponding to d-wave superconductor?' The magnetic analog of d-wave superconductor is the phase of an altermagnet (AM), which forms the basis of this work. d-wave superconductors have the superconducting term proportional to $(p_x^2-p_y^2)\tau_x$ in the Hamiltonian, where $\tau_x$ is a Pauli spin matrix that acts on the particle-hole sector. Similarly, if a term that is responsible for TRS breaking in the ferromagnetism is proportional to $(p_x^2-p_y^2)\si_z$ ($\si_z$ is a Pauli spin matrix that acts on  spin sector), a new class of magnetic phase known as altermagnetic phase~\cite{smejkal22a,smejkal22b,smejkal22c} shows up. 
Andreev reflection, Josephson effect, thermal transport, finite momentum Cooper-pairing in mesoscopic systems comprising altermagnets have been studied~\cite{papaj,sun23,ouass,zhou23,zhang23}. 

\begin{figure}[htb]
 \includegraphics[width=8cm]{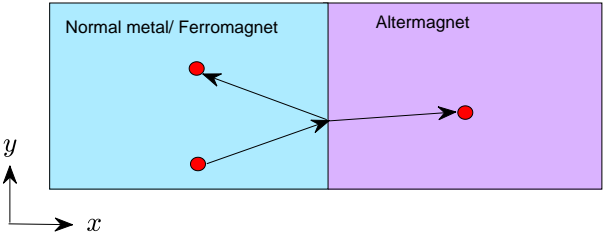}
 \caption{Schematic of junction between a normal metal (ferromagnet) and an altermagnet, depicting electron scattering event at the junction. }\label{fig:schem}
\end{figure}

In this work, we  study junctions of altermagnet with (i) normal metal and (ii) ferromagnet, depicted in Fig.~\ref{fig:schem} via continuum and lattice models. We map the parameters in the continuum model to parameters of the lattice model.  We calculate the conductivities of the junctions, and observe a finite spin polarized current at the junction between altermagnet and normal metal under an applied bias. The conductivity of the junction between altermagnet and ferromagnet depends on the direction and magnitude of the spin polarization in the ferromagnet. Our model provides a way to quantify  magnetoresistance of the ferromagnet-altermagnet junction. Thus, in addition to providing continuum description of junction of altermagnets with normal metals, we present a way to calculate physical properties that are pertinent to experimentally measurable quantities.

\section{Calculations and Results}
Altermagnets can be described by  the Hamiltonian $H_{\vec k} = -2t_0(\cos{k_xa}+\cos{k_ya})\si_0 + 2t_J(\cos{k_xa}-\cos{k_ya})\si_z$ in momentum-space, where $a$ is the lattice spacing and $\si_0$ is the identity matrix. In the limit where $t_J=0$ and $t_0\neq 0$, this Hamiltonian describes a normal metal. The term proportional to $t_J$ is responsible for an altermagnetic phase. For the case of a small nonzero $t_J$ ($0<t_J\ll t_0$), AM is in weak phase and in the limit $0<t_0\ll t_J$, AM is in strong phase. In the weak phase, the Fermi surfaces for up-spin and down-spin cross each other while in the strong phase the two Fermi surfaces are well separated as can be seen in Fig.~\ref{fig:FS} and Fig.~\ref{fig:almdisp}(a). Transition from the weak AM phase to strong AM phase occurs at $t_0=t_J$. 
\subsection{$t_0=0$, $t_J>0$}
To begin with, we take the limit $t_0=0$, $t_J>0$ since the term responsible for the altermagnetic phase dominates in this case~\cite{smejkal22a}. 
\subsubsection{Normal metal  - altermagnet junction: Lattice model}
 The Hamiltonian for a junction between a normal metal (NM) and AM can be written as: 
\bea 
H &=& \sum_{n_x=-\infty}^{0}\sum_{n_y=-\infty}^{\infty}[-t(c^{\dagger}_{n_x-1,n_y}c_{n_x,n_y}+h.c.) \nn \\ && -t(c^{\dagger}_{n_x,n_y+1}c_{n_x,n_y}+h.c.)   -\mu_{nl}c^{\dagger}_{n_x,n_y}c_{n_x,n_y}] \nn \\ &&- t' \sum_{n_y=-\infty}^{\infty} (c^{\dagger}_{0,n_y}c_{1,n_y} + h.c.) \nn \\ 
&& +  \sum_{n_x=1}^{\infty}\sum_{n_y=-\infty}^{\infty}[t_J(c^{\dagger}_{n_x+1,n_y}\si_zc_{n_x,n_y}+h.c.) \nn \\ && -t_J(c^{\dagger}_{n_x,n_y+1}\si_zc_{n_x,n_y}+h.c.)   -\mu_{al}c^{\dagger}_{n_x,n_y}c_{n_x,n_y}], ~~~~\label{eq:H}
\eea
where $c_{n_x,n_y}=[c_{n_x,n_y,\ua}, c_{n_x,n_y,\da}]^T$, $c_{n_x,n_y,\si}$ annihilates an electron of spin $\si$ at site $(n_x, n_y)$, $t$ is the hopping strength in the normal metal, $t_J$ is the strength of the spin- and direction- dependent hopping in the altermagnet, $\mu_{nl}$ and $\mu_{al}$ are chemical potentials of NM and AM respectively, $t'$ is the hopping strength on  bonds that connect NM to AM. The Pauli spin matrix $\si_z$ commutes with the Hamiltonian, and hence the eigenstates of $\si_z$ are also eigenstates of the Hamiltonian.  The dispersion on NM side is $E(\vec k)=-2t(\cos{k_xa}+\cos{k_ya})-\mu_{nl}$ for both  spins, while the dispersion on AM side is $E(\vec k) =2t_J(\cos{k_xa}-\cos{k_ya})-\mu_{al}$ for $\ua$-spin and $E(\vec k)=-2t_J(\cos{k_xa}-\cos{k_ya})-\mu_{al}$ for $\da$-spin.

\begin{figure}[htb]
 \includegraphics[width=8.6cm]{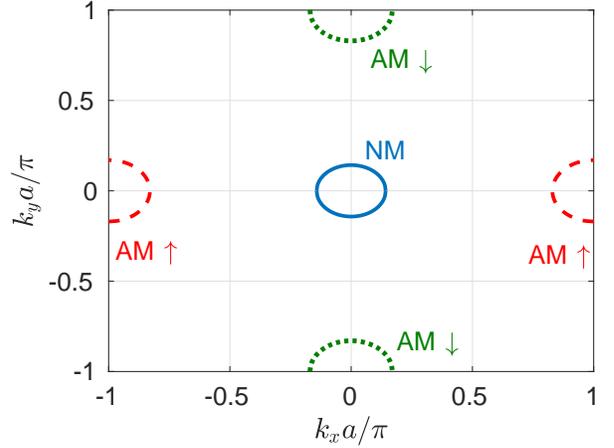}
 \caption{Fermi surfaces of normal metal (NM) and altermagnet (AM) near the band bottom. $t_J>t_0=0$ and AM is in the strong phase here. }\label{fig:FS}
\end{figure}

The NM dispersion near the band-bottom (which is at $k_x=k_y=0$),  can be written as $E(\vec k)=-4t-\mu_{nl}+t(k_x^2a^2+k_y^2a^2)$. In AM, for up-spin, the bandbottom is at $k_xa=\pm\pi, ~k_y=0$ and near the bandbottom,  dispersion can be written as $E(\vec k)=-4t_J-\mu_{al}+t_J[(k_xa\mp\pi)^2+k_y^2a^2]$ whereas for the down-spin, bandbottom is at $k_x=0, ~k_ya=\pm\pi$. Near the bandbottom, dispersion can be written as $E(\vec k)=-4t_J-\mu_{al}+t_J[k_x^2a^2+(k_ya\mp\pi)^2]$. We align the band bottoms of NM and AM by choosing $(-4t-\mu_{nl}) = (-4t_J-\mu_{al})$ and work at energies close to the bandbottom so that  quadratic approximation for dispersions on the two sides is valid. In Fig.~\ref{fig:FS},  Fermi surfaces of normal metal and altermagnet near the bandbottom are plotted. 
In scattering across the junction between NM and AM, with the junction located at bond $n_x=0$ to $n_x=1$ along $x$-direction, $k_y$ is taken to be the same on either sides since the system is translationally invariant along $y$-direction. While for up-spin electrons incident from NM, $k_y$ can be matched on AM side, for the down-spin electrons there is no $k_y$ on AM side that can be matched with $k_y$ of the electrons incident from NM. This implies that no current is transmitted into AM in the down-spin channel for energies near the bandbottom. 

We choose $\mu_{nl}=4t$ and $\mu_{al}=4t_J$ so that the band bottoms of NM and AM are aligned at $E=0$. 
We study transport at positive energies, small compared to the least of $t$ and $t_J$. 
The scattering wavefunction for an up-spin electron incident on the junction at energy $E>0$ making an angle $\th$ with the $x$-axis has the form $|\psi\ra=\sum_{n_x,n_y}\psi_{n_x,n_y}|n_x,n_y\ra$ where
\bea 
\psi_{n_x,n_y} &=& (e^{ik_xan_x} + r_{k,l} e^{-i k_xa n_x})e^{ik_yan_y}, ~~{\rm for~~}n_x \le 0, \nn \\ 
&=& t_{k,l}e^{i(k'_xan_x+k_yan_y)}, ~~{\rm for ~~}n_x\ge 1 ~. 
\eea
Here, $k_xa=\sqrt{E/t}\cos{\th}$, $k_ya=\sqrt{E/t}\sin{\th}$ and $k'_xa=\pi+\sqrt{E/t_J-k_y^2a^2}$. The scattering amplitudes $r_{k,l}$ and $t_{k,l}$ can be found using Schr\"odinger equation to be 
\bea 
r_{k,l} = -\f{t'^2e^{ik'_xa}+tt_Je^{ik_xa}}{t'^2e^{ik'_xa}+tt_Je^{-ik_xa}}, && 
t_{k,l} = \f{2itt'\sin{k_xa}}{t'^2e^{ik'_xa}+tt_Je^{-ik_xa}}.~~~ \label{eq:rktk-l}
\eea 
For values of $|\th|$ above a critical value, $k_ya$ can be large so that $k'_x$ is complex. In this case, the wave is not a plane wave on AM side and it carries no current. Also, we can see from eq.~\eqref{eq:rktk-l}, that $e^{ik'_xa}$ is real and $|r_{k,l}|=1$ resulting in perfect reflection. 

\subsubsection{Normal metal  - altermagnet junction: Continuum model}

Next, we formulate the scattering problem in the continuum language. Near the bandbottom, continuum Hamiltonian for NM and AM in the up-spin sector are given by 
\bea 
H &=& -t\big(\f{\Do^2}{\Do x^2}+\f{\Do^2}{\Do y^2}\big), ~~~{\rm for ~~} x<0 \nn \\ 
  &=& -t_J\big(\f{\Do}{\Do x} -i\pi \big)^2 - t_J \f{\Do^2}{\Do y^2}, ~~~{\rm for ~~} x>0 .\label{eq:Ham-cont} 
\eea 
On AM side, we have expanded the Hamiltonian around $k_xa=\pi$. 
The wavefunction for an up-spin electron incident at energy $E$ and angle $\th$ has the form 
\bea 
\psi(x,y) &=& (e^{ik_x x}+r_{k,c}e^{-ik_x x})e^{ik_yy}, ~~~{\rm for ~~} x<0 \nn \\ 
&=& t_{k,c}e^{i(k'_xx+k_yy)}, ~~~{\rm for ~~}x>0,  \label{eq:psic}
\eea
where $k_x,~ k'_x,~ k_y$ are same as earlier, and the scattering amplitudes $r_{k,c}, ~t_{k,c}$ need to be determined using the boundary conditions. 

 The $x$-components of the probability currents on NM and AM sides are given by $J_{NM}=2t~{\rm Im}[\psi^{\dag}\Do_x\psi]/\hbar$ and $J_{AM}=2t_J~{\rm Im}[\psi^{\dag}\Do_x\psi]/\hbar-2\pi t_J\psi^{\dag}\psi/\hbar$ respectively. Similar to the case of junction between quantum wires~\cite{soori23scat}, conservation of current at $x=0$ dictates the boundary conditions for scattering problem: 
 \bea 
 \psi(0^+) &=& c~\psi(0^-),~~ \nn \\  (\Do_x\psi-i\pi\psi)|_{0^+} &=& \big(\f{t}{c~t_J}\Do_x\psi-q\psi\big)_{0^-}, \label{eq:bc}
 \eea
where $c, q$ are two real parameters which describe the junction. 
 Using these boundary conditions, the scattering amplitudes can be shown to be: 
 \bea 
 r_{k,c} &=&  \f{t k_xa-t_Jc^2(k'_xa-\pi)-iqc^2t_J}{t k_xa+t_Jc^2(k'_xa-\pi)+iqc^2t_J} \nn \\ 
 t_{k,c} &=& \f{2tck_xa}{t k_xa+t_Jc^2(k'_xa-\pi)+iqc^2t_J} \label{eq:rktk-c}
 \eea 
 \subsubsection{Mapping the lattice model to the continuum model}
 In the limit of small $k_xa, ~(k'_xa-\pi)$, the expressions for scattering amplitudes in the lattice model [eq.~\eqref{eq:rktk-l}] and in the continuum model [eq.~\eqref{eq:rktk-c}] are the same for the choice of $c,~q$ given by $c=-t'/t_J$ and $q=(tt_J/t'^2-1)$. Thus, we have mapped the lattice model of the junction to the continuum model. 
  In such a scattering problem, there exists a critical angle of incidence beyond which $k'_x$ is complex and the wavefunction on AM side does not carry a current beyond the critical angle of incidence. Such a critical angle is given by $\th_C=\sin^{-1}[{\rm min}(\sqrt{t/t_J},1)]$. 
  
 The differential conductivity $G$ at bias $V$ which is the ratio between the differential change in current density $dI$ to the differential change in voltage applied $dV$ when the bias changes from $V$ to $V+dV$ is given by~\cite{suri21,soori2021} 
 \bea
 G(V) &=& \f{e^{2}}{h}\f{1}{2\pi}\sqrt{\f{eV}{t}}\int_{-\th_C}^{\th_C}d\th \cos{\th}(1-|r_k|^2),  \label{eq:cond}
 \eea
 where $r_k$ can be calculated from continuum model [$r_{k,c}$ from eq.~\eqref{eq:rktk-c}] or from the lattice model [$r_{k,l}$ from eq.~\eqref{eq:rktk-l}]. The current on NM side is proportional to  $(1-|r_k|^2)$, and the multiplicative factor  $\cos{\th}$ is due to the angle of incidence $\th$. 
 
 \subsubsection{Ferromagnet-altermagnet junction }
 Since the current is carried by electrons of up-spin alone, an applied bias results in charge current as well as spin current. The conductivity of a junction between  ferromagnet  and  altermagnet,  can probe the  dependence on  spin polarization. In junctions of AM  with ferromagnets, a  purely up-spin polarised ferromagnet will show a finite conductivity, while a purely down spin polarised ferromagnet will result in zero conductivity. But most ferromagnets have both up-spin and down-spin electrons, though to a different extent. We model the ferromagnet with Stoner model by adding a term proportional to $\si_z$ to the free electron model~\cite{stoner38}.
 The Hamiltonian describing ferromagnet-altermagnet junction can be written as 
\bea 
H &=& \Big[-t\big(\f{\Do^2}{\Do x^2}+\f{\Do^2}{\Do y^2}\big)-\mu\Big]\si_0-b\si_z, ~~~{\rm for ~~} x<0 \nn \\ 
  &=& \Big[-t_J\big(\f{\Do}{\Do x} -i\pi \big)^2 - t_J \f{\Do^2}{\Do y^2}\Big]\si_z, ~~~{\rm for ~~} x>0, \label{eq:Ham-fmam} 
\eea 
 where $b$ quantifies the extent of spin polarization in the ferromagnet and is called Stoner parameter. 

  \begin{figure}[htb]
  \includegraphics[width=8cm]{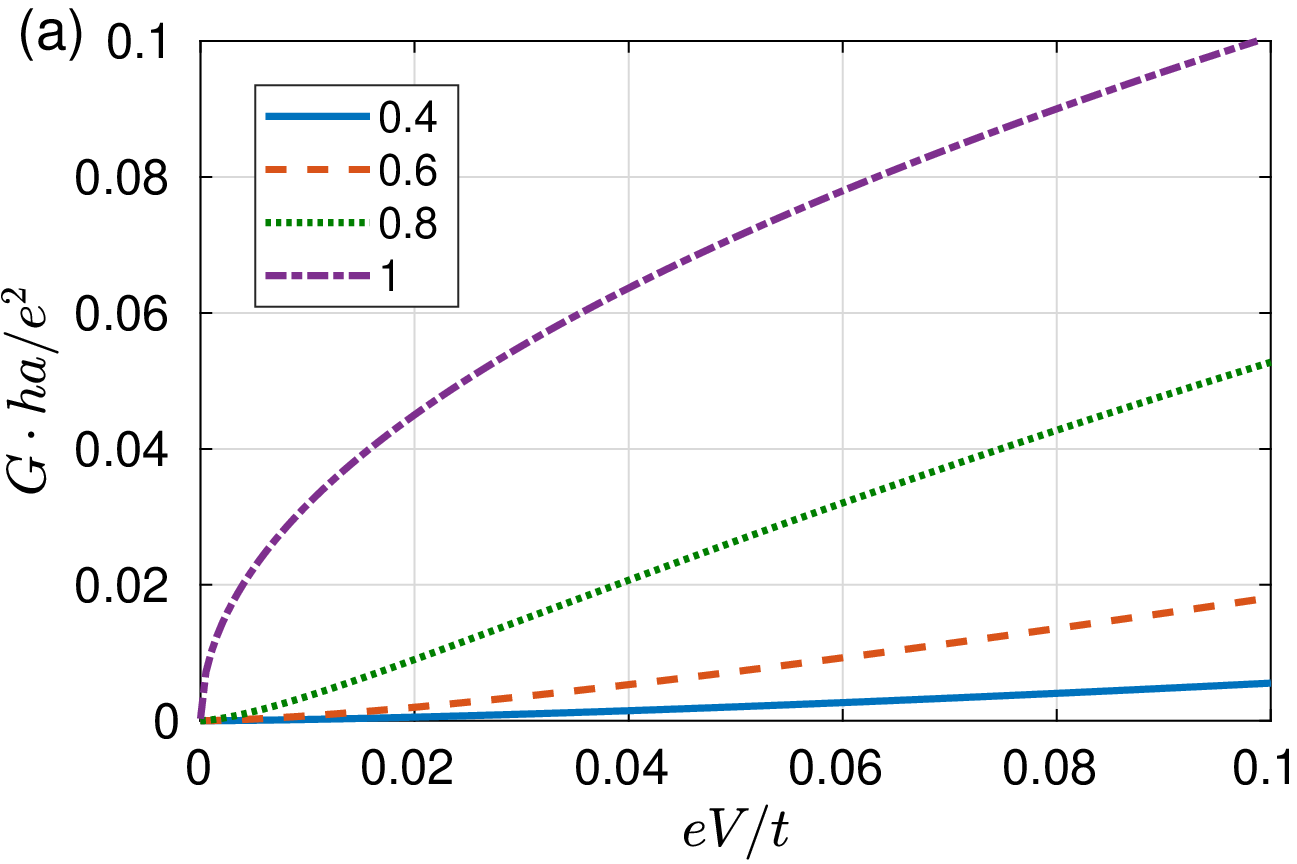}
  \includegraphics[width=8cm]{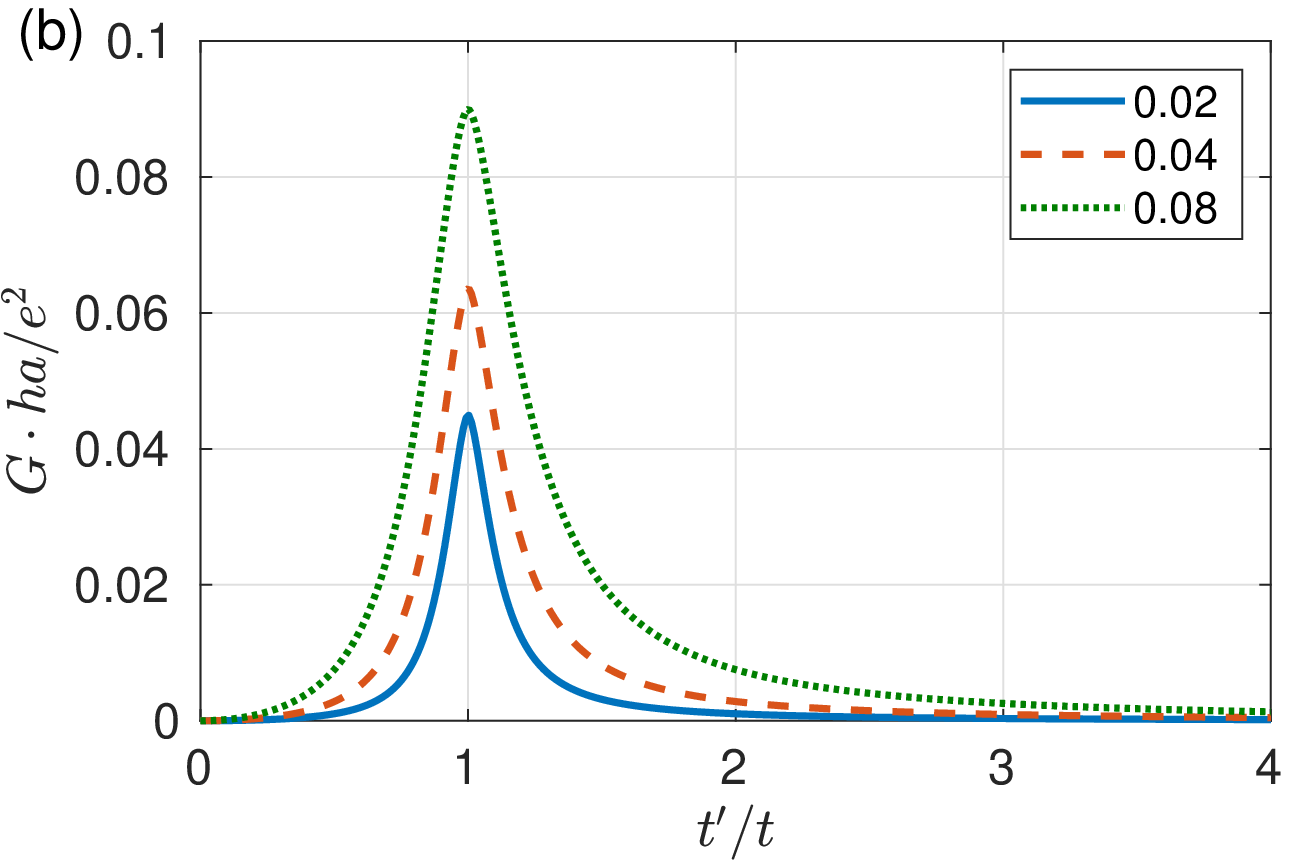}
  \caption{Conductivity of NM-AM junction versus (a) bias for different values of t'/t as indicated in the legend, (b) t'/t for different values of eV/t as indicated in the legend. Parameter: $t_J=t$. }\label{fig:G}
 \end{figure}

 We assume $b,\mu \ll t$ so that the band bottoms of the altermagnet and ferromagnet are close-by and  quadratic approximation for dispersion is valid. Similar to the case of normal metal - altermagnet junction, transport occurs only in the up-spin sector here. The scattering wavefunction for an up-spin electron incident at energy $E>0$, at an angle of incidence $\th$ takes the same form as in eq.~\eqref{eq:psic}, except that $k=\sqrt{(E+\mu+b)/t}$ and $k_xa=ka\cos{\th}$. Also $\th_C$ is now given by $\th_C=\sin^{-1}[{\rm min}(1,\sqrt{t/t_J-(\mu+b)/t_Jk^2a^2})]$. The boundary conditions in eq.~\eqref{eq:bc} can be used to calculate the scattering coefficients. The scattering coefficients take the same form as in eq.~\eqref{eq:rktk-c}, except that $k_x$ is given by $k_xa=\sqrt{(E+\mu+b)/t}\cos{\th}$ and the formula for conductivity will change to 
 \bea 
 G(V) &=&  \f{e^{2}}{h}\f{1}{2\pi}\sqrt{\f{eV+\mu+b}{t}}\int_{-\th_C}^{\th_C}d\th \cos{\th}(1-|r_k|^2).~~~~  
 \eea
 In spintronics, the dependence of the conductivity on spin polarization of the ferromagnet is quantified by magnetoresistance which is defined as $MR = [\rho(b)-\rho(0)]100/\rho(0)$, where $\rho(b)=1/G(b)$, and $G(b)$ is the differential conductivity evaluated for the junction with Stoner parameter $b$.

\subsubsection{Results}

 \begin{figure}[htb]
  \includegraphics[width=8cm]{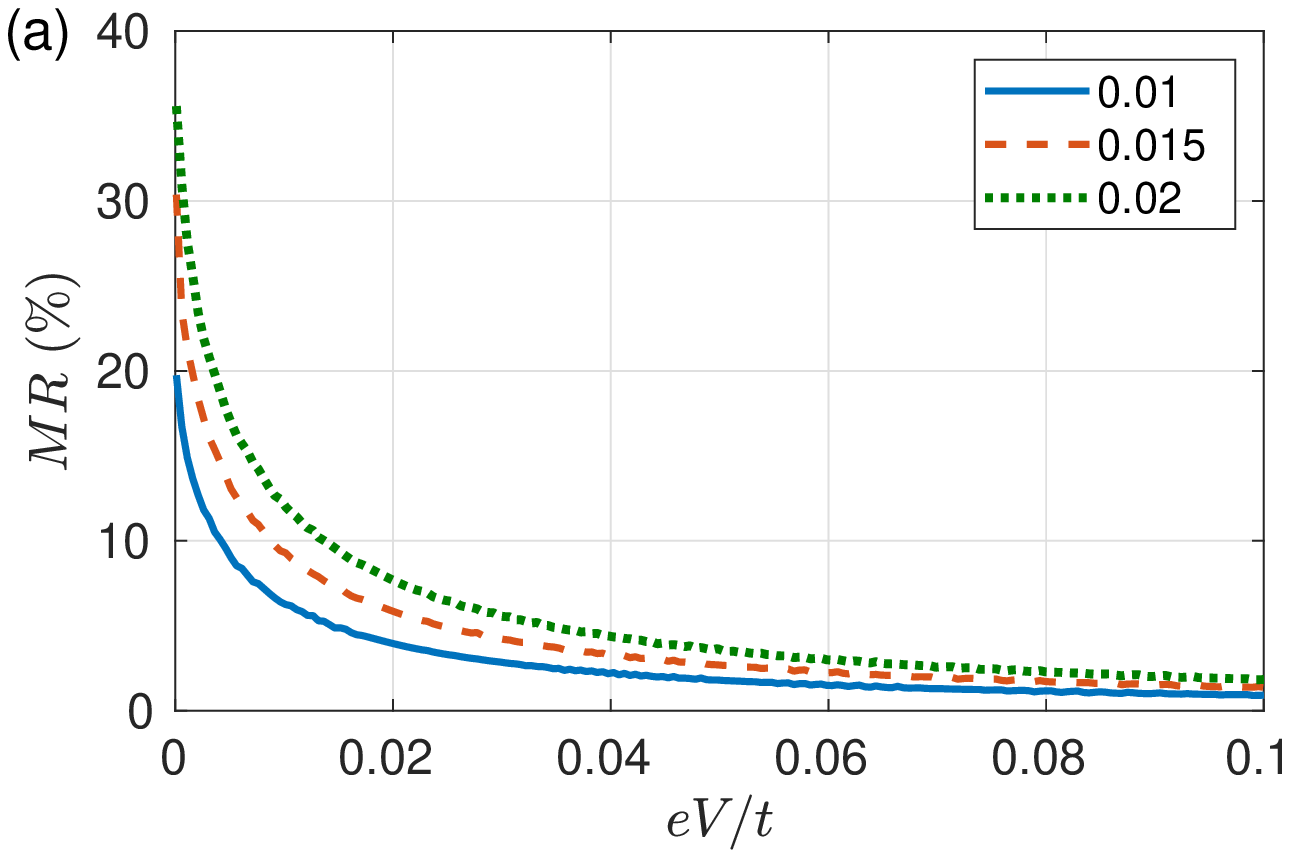}
  \includegraphics[width=8cm]{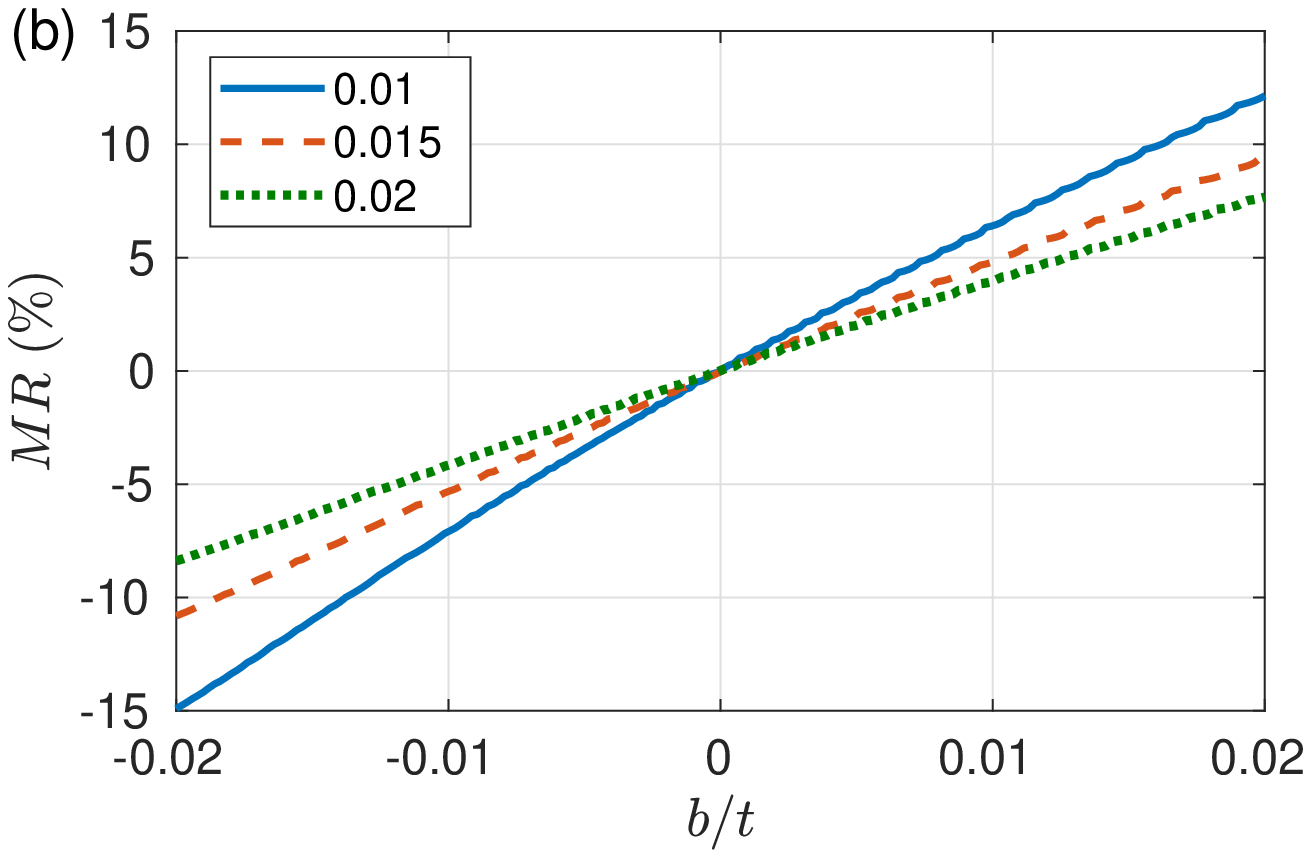}
  \caption{Magnetoresistance of the ferromagnet-altermagnet
junction as a function of: (a) bias for different values of  $b/t$ as shown in the legend (b) Stoner parameter $b$ for different values of $eV/t$ as shown in the legend.  Parameters: $t_J=t$, $t'=t$, $\mu=0.02t$.}\label{fig:MR}
 \end{figure}
 
 We numerically calculate the conductivity of NM-AM junction and plot it as a function of the voltage bias $V$ in Fig.~\ref{fig:G}(a) for the choice of parameters $t_J=t$, for different values of $t'/t$ indicated in the legend. In Fig.~\ref{fig:G}(b), we plot the differential conductivity as a function of $t'/t$ for different values of the bias $eV/t$ shown in the legend. We find that differential conductivity increases with the bias voltage. This is because of the density of states in NM and AM regions, which increases as the energy  is increased. The conductivity is peaked at $t'=t$. This is for the case $t_J=t$, which is similar to the conductance of a junction between two quantum wires, which is maximum for $t'=t$~~\cite{soori23scat}. The location of the peak in $G$ versus $t'/t$ plot depends on the value of $t_J/t$.

 For the ferromagnet-altermagnet junction, we plot the magnetoresistance  as a function of the bias in Fig.~\ref{fig:MR}~(a) for different values of $b/t$ indicated in the legend, choosing $t_J=t'=t$ and $\mu=0.02t$. MR decreases with the bias. This is because, at higher energies, the density of states is higher and the effect of nonzero $b$ (Stoner parameter) is reduced at higher values of bias. In Fig.~\ref{fig:MR}~(b), the magnetoresistance is plotted as a function of $b/t$ for different values of $eV/t$ shown in the legend. With increasing spin polarization in the ferromagnet, MR of the junction increases. 
 
 \subsection{$t_0>t_J$}
 In this subsection, we will discuss the case $t_0>t_J>0$. 
 In this case, the bandbottoms for both  spins will be at $k_x=k_y=0$. However, the Fermi surfaces near the bandbottom become anisotropic and the Fermi surfaces for the two spins do not overlap. This leads to different transmission probabilities for the two spins,  and hence a spin current in  response to bias in a NM-AM junction.
   The Hamiltonian for a junction between NM and AM can be written as: 
\bea 
H &=& \sum_{n_x=-\infty}^{0}\sum_{n_y=-\infty}^{\infty}[-t(c^{\dagger}_{n_x-1,n_y}c_{n_x,n_y}+h.c.)  \nn \\ && -t(c^{\dagger}_{n_x,n_y+1}c_{n_x,n_y}+h.c.)   -\mu_{nl}c^{\dagger}_{n_x,n_y}c_{n_x,n_y}] \nn \\ &&- t' \sum_{n_y=-\infty}^{\infty} (c^{\dagger}_{0,n_y}c_{1,n_y} + h.c.) \nn \\ 
&& +  \sum_{n_x=1}^{\infty}\sum_{n_y=-\infty}^{\infty}[\{c^{\dagger}_{n_x+1,n_y}(t_J\si_z-t_0\si_0)c_{n_x,n_y}+h.c.\}  \nn \\ && -\{c^{\dagger}_{n_x,n_y+1}(t_J\si_z+t_0\si_0)c_{n_x,n_y}+h.c.\}  \nn \\ &&  -\mu_{al}c^{\dagger}_{n_x,n_y}c_{n_x,n_y}], ~~~~\label{eq:H2}
\eea
where $c_{n_x,n_y}=[c_{n_x,n_y,\ua}, c_{n_x,n_y,\da}]^T$, $c_{n_x,n_y,\si}$ annihilates an electron of spin $\si$ at site $(n_x, n_y)$, $t_0$ is the hopping strength in the altermagnet and other symbols have same meaning as mentioned  earlier in the context of eq.~\eqref{eq:H}.
The Pauli spin matrix $\si_z$ commutes with the Hamiltonian, and hence the eigenstates of $\si_z$ are also eigenstates of the Hamiltonian.  The dispersion on NM side is $E(\vec k)=-2t(\cos{k_xa}+\cos{k_ya})-\mu_{nl}$ for both spins, while the dispersion on AM side is $E(\vec k)=-2(t_0-t_J)\cos{k_xa}-2(t_0+t_J)\cos{k_ya}-\mu_{al}$ for $\ua$-spin and   $E(\vec k)=-2(t_0+t_J)\cos{k_xa}-2(t_0-t_J)\cos{k_ya}-\mu_{al}$ for $\da$-spin.  
For $t_0>t_J>0$, the band bottoms on AM side for both  spins is at $(k_x,k_y)=(0,0)$.

\begin{figure}[htb]
 \includegraphics[width=8.01cm]{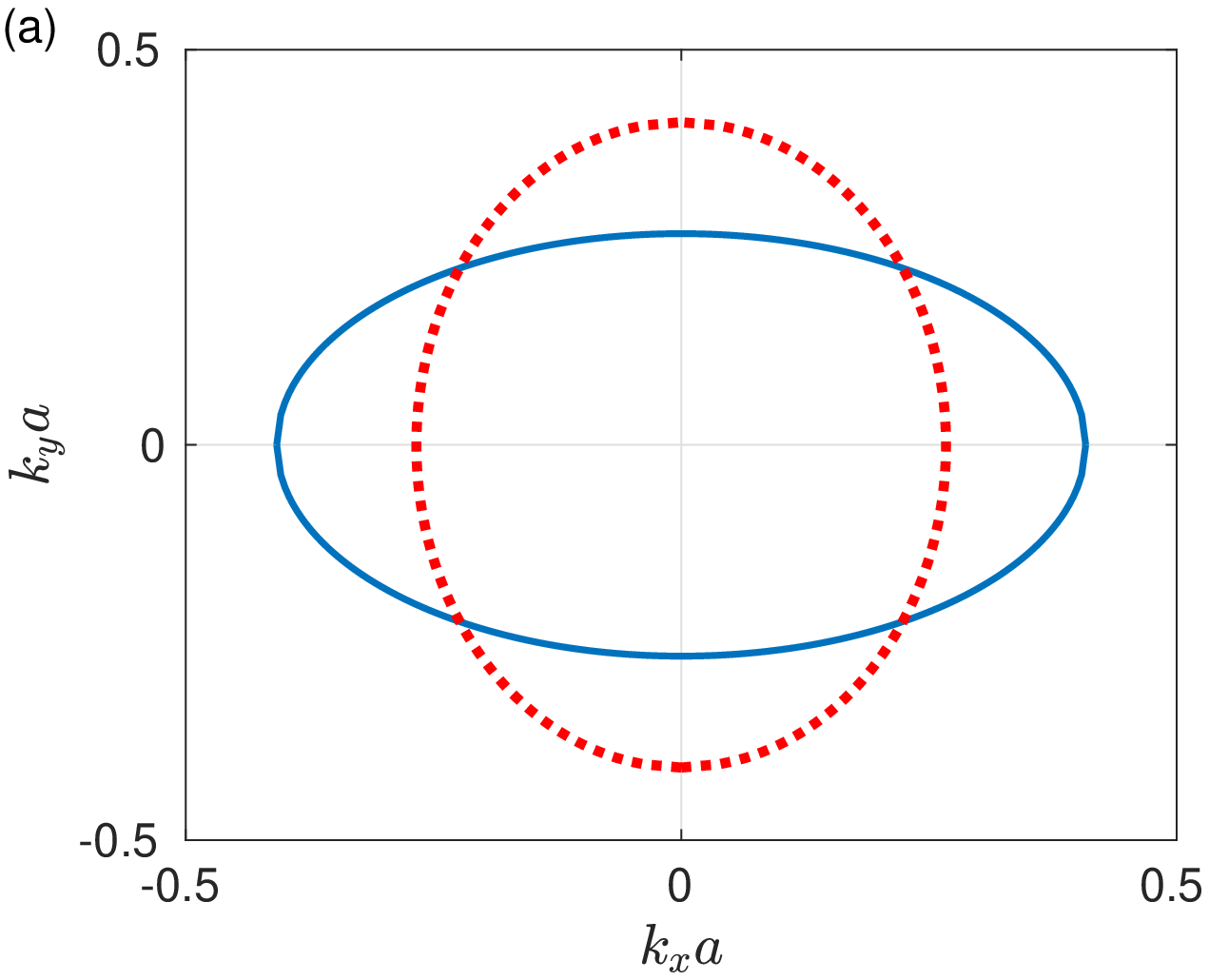}
 \includegraphics[width=8cm]{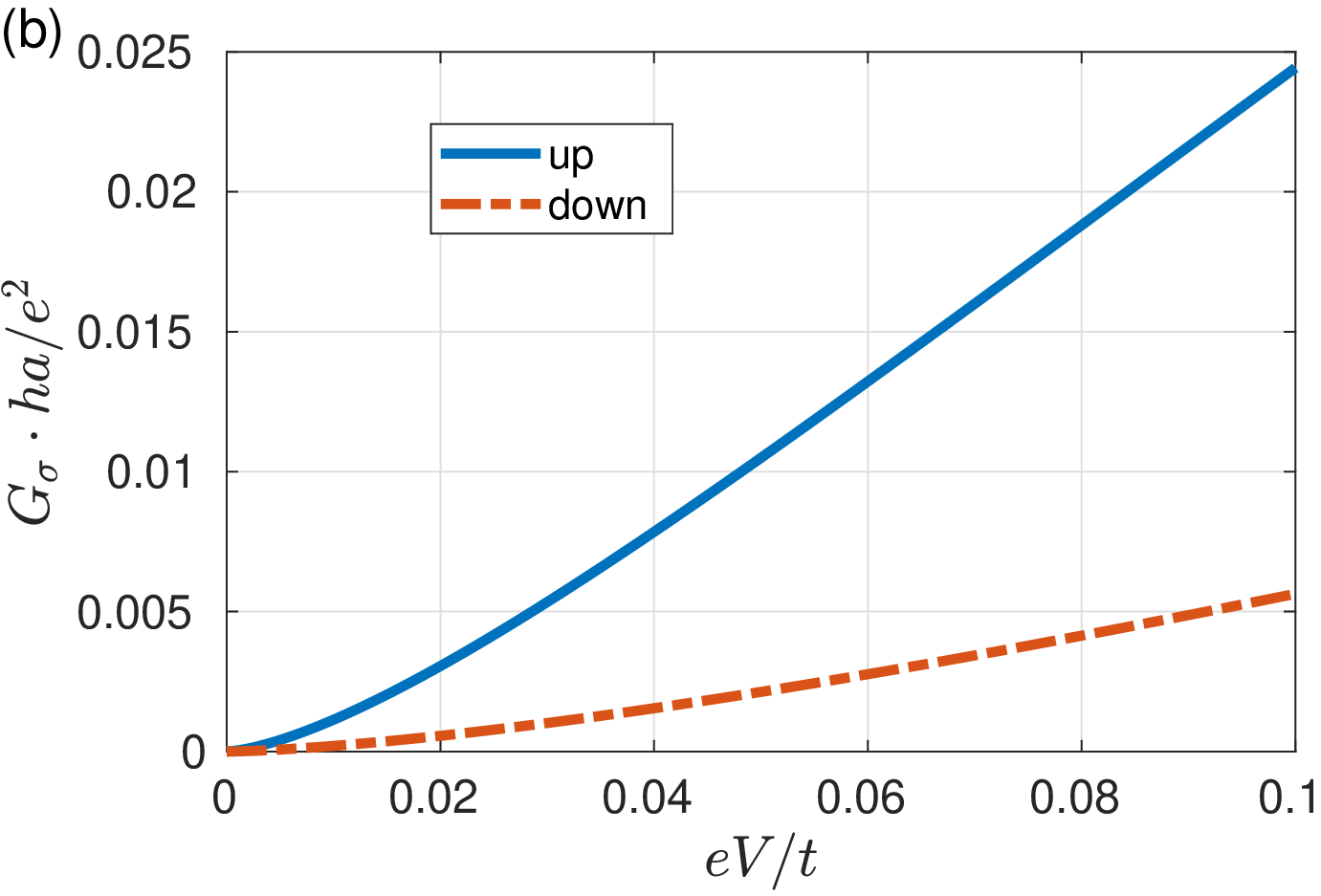}
 \caption{(a) Fermi surfaces for the up-spin (solid line) and down spin (dotted line) on altermagnet at Fermi energy $E=0.1t$, $t_0=t$, $t_J=0.4t$. The AM is in the weak phase here.  (b) Conductivity  due to spin-$\si$ bands versus bias for $t_0=t$, $t_J=0.4t$ and $t'=0.5t$.}\label{fig:almdisp}
\end{figure}

We choose $\mu_{nl}=4t$ and $\mu_{al}=4t_0$ so that the band bottoms of NM and AM are aligned at $E=0$. In the limit $t_0>t_J$, for small energies, dispersions for the $\ua/\da$ bands in the altermagnet are $E_\si(\vec k)=(t_0-s_{\si}t_J)(k_xa)^2+(t_0+s_{\si}t_J)(k_ya)^2$, where  $s_{\ua}=+$, $s_{\da}=-$. In Fig.~\ref{fig:almdisp}(a), the Fermi surfaces for the two spin bands on altermagnet is depicted. 

We study transport at positive energies, close to the band bottom. 
The scattering wavefunction for a $\si$-spin electron incident on the junction at energy $E>0$ making an angle $\th$ with the $x$-axis has the form $|\psi_{\si}\ra=\sum_{n_x,n_y}\psi_{n_x,n_y,\si}|n_x,n_y,\si \ra$ where
\bea 
\psi_{n_x,n_y,\si} &=& (e^{ik_xan_x} + r_{k,\si} e^{-i k_xa n_x})e^{ik_yan_y}, ~~{\rm for~~}n_x \le 0, \nn \\ 
&=& t_{k,\si}e^{i(k'_{x,\si}an_x+k_yan_y)}, ~~{\rm for ~~}n_x\ge 1 ~. 
\eea
Here, $k_xa=\sqrt{E/t}\cos{\th}$, $k_ya=\sqrt{E/t}\sin{\th}$ and $k'_{x,\si}a=\sqrt{[E-(t_0+s_{\si} t_J)k_ya^2]/[t_0-s_{\si}t_J]}$. The scattering amplitudes $r_{k,\si}$ and $t_{k,\si}$ can be found using Schr\"odinger equation to be 
\bea 
r_{k,\si} &=& -\f{t(t_0-s_{\si}t_J)e^{ik_xa}-t'^2e^{ik'_{x,\si}a}}{t(t_0-s_{\si}t_J)e^{-ik_xa}-t'^2e^{ik'_{x,\si}a}}, {~~\rm and~~} \nn \\ 
t_{k,\si} &=& \f{-2itt'\sin{k_xa}}{t(t_0-s_{\si}t_J)e^{-ik_xa}-t'^2e^{ik'_{x,\si}a}} \label{eq:rktk}
\eea 

The differential conductivity $G_{\si}(V)$ due to electrons of spin $\si$ can be obtained using the formula similar to eq.~\eqref{eq:cond}, except that $r_{k}$ there is replaced with $r_{k,\si}$. Also, the range of integration in eq.~\eqref{eq:cond} can be taken to be $(-\pi/2,\pi/2)$ since the $\th_C$ will change for this case and for $|\th|>\th_C$, $|r_k|=1$.  
 Using eq.~\eqref{eq:rktk} in the formula for $G_{\si}(V)$, we numerically calculate the conductivities due to up/down-spin bands for $t_0=t$, $t_J=0.4t$, $t'=0.5t$ and plot in Fig.~\ref{fig:almdisp}(b). In this case, the Fermi surface of the normal metal is circular with  a radius $k_F$ which is between the major and minor axes of the elliptical Fermi surface of the altermagnet. Hence, for up-spin, all the states in the altermagnet carry current while for down-spin, states with $|k_y|>k_F$ do not carry current making the conductivity due to up spin larger compared to the conductivity due to down spin. 
 We clearly see that the conductivities due to the two spin components are different, and hence a spin current accompanies charge current when a bias is applied across the normal metal-altermagnet junction in the case $t_0>t_J>0$.

\section{Discussion}
 
 If  crystallographic orientation of the AM is rotated by $90^{\circ}$, conductivity of the  NM-AM junction will not change; instead, the current will be carried by electrons of opposite spin. Applying bias across a junction between  altermagnet and  normal metal generates a spin current. Therefore, the value of the spin current generated depends on the crystallographic orientation of the altermagnet. This can have applications in spin based memory devices. Purely spin polarised ferromagnet with up-spin electrons will show a finite conductivity, while the one with down-spin polarized electrons will show zero conductivity. Under $90^{\circ}$ rotation, a purely spin polarized ferromagnet with down-spin electrons will show a finite conductivity, while the one with up-spin electrons will show zero conductivity. The magnetoresistance plots of Fig.~\ref{fig:MR} will evolve so that $MR\to- MR$ under such rotation. Such a dependence of the magnetoresistance on the crystallographic orientation is a signature exclusive to altermagnetic phase and is absent in ferromagnetic or antiferromagnetic phases.

\section{Summary}
Altermagnetic metal has dominant altermagnetic order for $t_J\gg t_0$. First, we studied the case $t_0=0$ and $t_J\neq 0$ where the altermagnetic order is dominant. 
 We have studied a junction between  a normal metal and an altermagnet using lattice and continuum models. In the continuum model, we identified the boundary condition that obeys current conservation. We mapped the parameters in the lattice model to those of the continuum model. We calculated the conductivity of the junction which gets nonzero contribution from only one spin, depending on the crystallographic orientation of the altermagnet. Further, we studied a junction between ferromagnet and altermagnet. We find that the magnetoresistance of the junction depends on  spin polarization and the bias, in addition to crystallographic orientation of AM. 
 Second, we also studied the case $t_0>t_J>0$ where the altermagnetic order is weak. However, even in this case, we found that a junction of such AM with normal metal exhibits nonzero spin current when a bias is applied. 
 Our model provides a simple analytical approach to characterize a AM based system and also quantifies experimentally measurable physical quantities.

\acknowledgements
SD and AS thank DST-INSPIRE Faculty Award (Faculty Reg. No.~:~IFA17-PH190) and SERB Core Research grant (CRG/2022/004311) for financial support. AS thanks funding from University of Hyderabad Institute of Eminence PDF.  DS acknowledges funding from the European Union’s Horizon 2020 research and innovation programme under the Marie Sklodowska-Curie grant agreement No 899987.

\bibliography{ref_almag}

\end{document}